# Intrinsic transverse spin angular momentum of fiber eigenmodes

Liang Fang and Jian Wang*

*Wuhan National Laboratory for Optoelectronics, School of Optical and Electronic Information, Huazhong University of Science and Technology, Wuhan 430074, Hubei, China. *Correspodning author: jwang@hust.edu.cn*

We study the transverse spin angular momentum of fiber-guided eigenmodes. As an intrinsic property of fiber eigenmodes, it becomes considerable in optical nanofiber, and especially, increases sharply in the evanescent field. We derive analytical expressions of this angular momentum, and present its density distribution inside and outside nanofiber. Significantly, we find the optimal ratio of fiber core radius to the wavelength to obtain the maximum surface transverse spin. For instance, the optimal fiber radius is one-fourth of the wavelength for the fundamental mode. Furthermore, we investigate the spin flow of each fiber-guided mode on the surface of nanofiber and the transverse mechanical effects originated from the transverse spin on small particles.

Keywords: Transverse spin; photonic angular momentum; nanofibers; fiber-guided eigenmodes.

The angular momentum (AM) is regarded as one of the most significant properties of electromagnetic fields, divided into spin angular momentum (SAM) and orbital angular momentum (OAM) [1]. The inherent SAM of light is yielded by the helicity of electric vector field along the propagation direction characterized by the well-known circular polarization [2]. The extrinsic OAM is associated with beam trajectory dependent of transverse coordinates of the beam centroid, similar to mechanical AM of classics particles [3]. Nevertheless, the intrinsic OAM arises from the helical phase inside vortex beams or structured wave fields determined by the topological charge number [4]. Momentum exchange generally happens in an interaction between two objects in macroscopic view, as well as that between ray radiation and microcosmic particle due to particle-like nature of light [5]. The linear momentum and angular momentum are usually involved in this process, as mechanical consequences, which correspondingly produces radiation forces and torques on particles or atoms [6].

Besides these AM contents, there exists another kind of SAM for structured optical fields, the transverse SAM (TSAM), that was recently discovered by A. Aiello and K. Y. Bliokh et al. and has attracted a rapidly growing interest [7]. It can been found in two-wave interference and evanescent waves [8], or in focused Gaussian beam [9]. As a new type of AM of light, the TSAM exhibits unique features in sharp contrast to the usual longitudinal SAM [10], such as the effects of spin-momentum locking and lateral forces [11]. It is independent of the helicity and is transverse, orthogonal to propagation direction. Its dominant ability is to achieve spin-controlled unidirectional propagation of light in nanofiber, surface plasmon-polaritons, and photonic-crystal waveguide [12], connected with the quantum spin Hall effect of light [13]. It may be potentially applied to integrated optical devices for fiber-based spin-logic classical and quantum implementations [14].

In generally weakly guiding fiber, the quantum numbers of both the well-known longitudinal SAM and OAM of guided-modes take integers in paraxial approximation, in addition, there intrinsically exits TSAM, but it is very faint. It is ascribed to the small component of longitudinal electric field that has a phase difference of $\pi/2$ relative to the transverse field, but can be neglected. However, in high-contract index fibers, such as nanofiber and hollow ring-core fiber [15], the longitudinal electric field becomes large enough, so that the guided modes cannot be regarded as propagating in paraxial axis. Consequently, on one hand, it leads to non-integer spin and orbit quantum numbers due to the non-negligible spin-orbit coupling of AM [15, 16]. On the other hand, it brings about a transverse polarization ellipticity and thus a considerable TSAM. However, so far, there have been limited research efforts to comprehensively study the TSAM of fiber eigenmodes. In this article, we study in detail the TSAM of fiber eigenmodes. We focus on this TSAM in nanofibers formed by a cylindrical silica core and equivalent air cladding. The nanofiber as a subwavelength-diameter waveguide has an intense evanescent field, and has a wide range of potential applications, such as being designed as micro- and nano-photonic devices [17]. Here we start from the fundamental theories in optics, derive analytical expressions of the TSAM of each fiber-guided eigenmode. We show its exact density distribution and analyze the optimized ratio of fiber core radius to wavelength to obtain maximum surface TSAM on nanofibers. Additionally, we present the spin flow of each fiber mode on nanofiber surface, and calculate the transverse torque and lateral force originated from TSAM.

We first present the general expression of energy density, helicity and SAM density of a monochromatic optical field in dielectric waveguide in SI Unit rather than Gaussian unit, which fulfills the Maxwell equations. The time-averaged densities of energy per unit length is

$$W = \frac{1}{2}\left(\varepsilon\left|\mathbf{E}\right|^2 + \mu\left|\mathbf{H}\right|^2\right), \tag{1}$$

where $\varepsilon$ and $\mu$ indicate real permittivity and permeability, respectively, $\mathbf{E}$ and $\mathbf{H}$ stand for the spatial distributions of electric and magnetic vectorial field, respectively. The polarization ellipticities for each components are [7, 18]

$$\boldsymbol{\varphi}_E = -i\left(\mathbf{E}^* \times \mathbf{E}\right), \tag{2}$$

$$\boldsymbol{\varphi}_H = -i\left(\mathbf{H}^* \times \mathbf{H}\right). \tag{3}$$

The helicity density introduced by electric field is given by [19]

$$K = \frac{i}{2\omega c}\left(\mathbf{E}^* \cdot \mathbf{H}\right), \tag{4}$$

where $\omega$ is angular frequency of light and $c$ is light velocity in vacuum. It is associated with the characterization of helicity coupling between matter and light [20]. The SAM density of light introduced from both the electric and magnetic field parts is given by,

$$\mathbf{S} = \frac{1}{2\omega}\left(\varepsilon_0 \boldsymbol{\varphi}_E + \mu_0 \boldsymbol{\varphi}_H\right), \tag{5}$$

where $\varepsilon_0$ and $\mu_0$ stand for permittivity and permeability in vaccum, respectively. However, compared with the dominant interaction between electric field and natural particles, the magnetic field acts very weakly on them. Hence, we focus on the SAM originated from the electric field and its corresponding mechanical effect. The optical torque that acts on a spherical particle due to the transverse spin of light is written as

$$\mathbf{T} = \omega \operatorname{Im}\left(\alpha_{ee}\right)\mathbf{S}, \tag{6}$$

And the lateral force produced by the transverse spin density on chiral particle is given by

$$\mathbf{F}_s = \frac{c\omega k_0^4}{3\pi\varepsilon_0} \operatorname{Re}\left(\alpha_{ee}\alpha_{em}^*\right)\mathbf{S}_\perp, \tag{7}$$

where $k_0 = 2\pi/\lambda$ is wave number and wavelength of light in free-space with $\lambda$ being wavelength, $\alpha_{ee}$ is electric-electric polarizability of the particle, $\alpha_{em}$ indicates the chirality property of the material of the particle [21], and $\mathbf{S}_\perp$ denotes the transverse component of SAM.

In the cylindrical coordinate, the corresponding three components of SAM can be expressed as,

$$\mathbf{S}_r = -\frac{i\varepsilon_0}{2\omega}\left(E_\phi^* E_z - E_z^* E_\phi\right)\vec{\mathbf{e}}_r, \tag{8}$$

$$\mathbf{S}_\phi = -\frac{i\varepsilon_0}{2\omega}\left(E_z^* E_r - E_r^* E_z\right)\vec{\mathbf{e}}_\phi, \tag{9}$$

$$\mathbf{S}_z = -\frac{i\varepsilon_0}{2\omega}\left(E_r^* E_\phi - E_\phi^* E_r\right)\vec{\mathbf{e}}_z, \tag{10}$$

where $r$ and $\phi$ denote radial and azimuthal positions, respectively. $E_r$, $E_\phi$ and $E_z$ represent radial, azimuthal and longitudinal components of electric field, respectively.

We can get $\left|\mathbf{S}_z\right| = K = 0$, which means that the fiber eigenmodes do not have longitudinal SAM and helicity, but there exits transverse SAM more or less yielded by the non-zero transverse component of electric polarization ellipticity according to Eq. (2). As for the high-contrast-index fibers, such as nanofiber, there is naturally a large longitudinal electric field with a phase difference of $\pi/2$ relative to transverse field, giving rise to a considerable TSAM $\mathbf{S}_r$ and $\mathbf{S}_\phi$. Besides it, with restriction to the boundary conditions of monochromatic field, the radial component of evanescent field trailing outside nanofiber become so intense sharply that would generate outstanding transverse SAM.

There are three types of fiber eigenmodes, TM $\left(H_z = 0\right)$, TE $\left(E_z = 0\right)$, and hybrid modes HE/EH modes [22]. Except for the TE mode, any eigenmodes possess the longitudinal component of electric field and thus can form the intrinsic TSAM. It should be noted that in nanofiber, a high-contrast -index waveguide, the eigenmodes could not be degenerated. They can transmit along fibers as independent vector modes. The field distribution of TM mode with zero order of azimuthal factor within fiber core $\left(0 \le r < a\right)$ is expressed as

$$E_r = jA\beta \frac{a}{u} J_1\left(\frac{u}{a}r\right), \tag{11}$$

$$E_z = AJ_0\left(\frac{u}{a}r\right), \tag{12}$$

and that in fiber cladding or surrounding of nanofiber $\left(r \ge a\right)$ is given by

$$E_r = -jA\beta \frac{aJ_0\left(u\right)}{wK_0\left(w\right)} K_1\left(\frac{w}{a}r\right), \tag{13}$$

$$E_z = A\frac{J_0\left(u\right)}{K_0\left(w\right)} K_0\left(\frac{w}{a}r\right), \tag{14}$$

where $\beta$ is the longitudinal propagation constant of TM mode, $u = a\sqrt{n_1^2 k_0^2 - \beta^2}$ and $w = a\sqrt{\beta^2 - n_2^2 k_0^2}$ indicate the normalized transverse wave numbers, respectively, $n_1$ and $n_2$ denote the refractive index of fiber core and cladding, respectively, $a$ is the radius of fiber core. $J_n$ and $K_n$ stand for the $n$-th order Bessel functions of the first kind and the $n$-th order modified Bessel functions of the second kind, respectively, and $A$ denotes the amplitude coefficient dependent of fiber parameters and the total power of corresponding mode.

The field distribution of HE/EH modes with $n$-th order of azimuthal factor in fiber core $\left(0 \le r < a\right)$ is expressed as

$$E_r = -jA\beta \frac{a}{u}\left[\frac{\left(1-s_0\right)}{2} J_{n-1}\left(\frac{u}{a}r\right) - \frac{\left(1+s_0\right)}{2} J_{n+1}\left(\frac{u}{a}r\right)\right]\cos\left(n\phi + \varphi_0\right) \tag{15}$$

$$E_\phi = jA\beta \frac{a}{u}\left[\frac{\left(1-s_0\right)}{2} J_{n-1}\left(\frac{u}{a}r\right) + \frac{\left(1+s_0\right)}{2} J_{n+1}\left(\frac{u}{a}r\right)\right]\sin\left(n\phi + \varphi_0\right) \tag{16}$$

$$E_z = AJ_n\left(\frac{u}{a}r\right)\cos\left(n\phi + \varphi_0\right), \tag{17}$$

and that in fiber cladding or surrounding of nanofiber $\left(r \ge a\right)$ is given by

$$E_r = -jA\beta \frac{aJ_n\left(u\right)}{wK_n\left(w\right)}\begin{bmatrix}\frac{\left(1-s_0\right)}{2} K_{n-1}\left(\frac{w}{a}r\right) \\ + \frac{\left(1+s_0\right)}{2} K_{n+1}\left(\frac{w}{a}r\right)\end{bmatrix}\cos\left(n\phi + \varphi_0\right), \tag{18}$$

$$E_\phi = jA\beta \frac{aJ_n\left(u\right)}{wK_n\left(w\right)}\begin{bmatrix}\frac{\left(1-s_0\right)}{2} K_{n-1}\left(\frac{w}{a}r\right) \\ - \frac{\left(1+s_0\right)}{2} K_{n+1}\left(\frac{w}{a}r\right)\end{bmatrix}\sin\left(n\phi + \varphi_0\right), \tag{19}$$

$$E_z = A\frac{aJ_n\left(u\right)}{wK_n\left(w\right)} K_n\left(\frac{w}{a}r\right)\cos\left(n\phi + \varphi_0\right), \tag{20}$$

and

$$s_0 = \frac{n\left(\frac{1}{u^2} + \frac{1}{w^2}\right)}{\left[\frac{J_n'(u)}{uJ_n(u)} + \frac{K_n'(w)}{wK_n(w)}\right]}, \tag{21}$$

where $\varphi_0$ is the initial phase.

From Eq. (12) and (14), one can see that the TM mode has a large longitudinal electric field, especially at the center of the field distribution shown Fig. 1(a), which can be used for electron acceleration, optical trapping, and laser machining, etc. [23]. As for HE/EH modes in weakly guiding fibers, the longitudinal electric field is so tiny that the TSAM can be ignored, whereas in high-contrast-index fibers, this quantity would become considerable because of the large longitudinal electric field as shown in Fig. 2(a). Inserting Eqs. (11)-(14) and (15)-(20) to Eqs. (9) and (10), one can get the density distribution of TSAM per unit length as follows.

For $TM_{01}$ mode, only exits the azimuthal component, i.e.,

$$S_\phi = \begin{cases} \frac{A^2 a\beta\varepsilon_0}{2\omega u} J_0\left(\frac{u}{a}r\right) J_1\left(\frac{u}{a}r\right) & r < a \\ -\frac{A^2 a\beta\varepsilon_0 J_0^2(u)}{2\omega w K_0^2(w)} K_0\left(\frac{w}{a}r\right) K_1\left(\frac{w}{a}r\right) & r \geq a \cdot \end{cases} \tag{22}$$

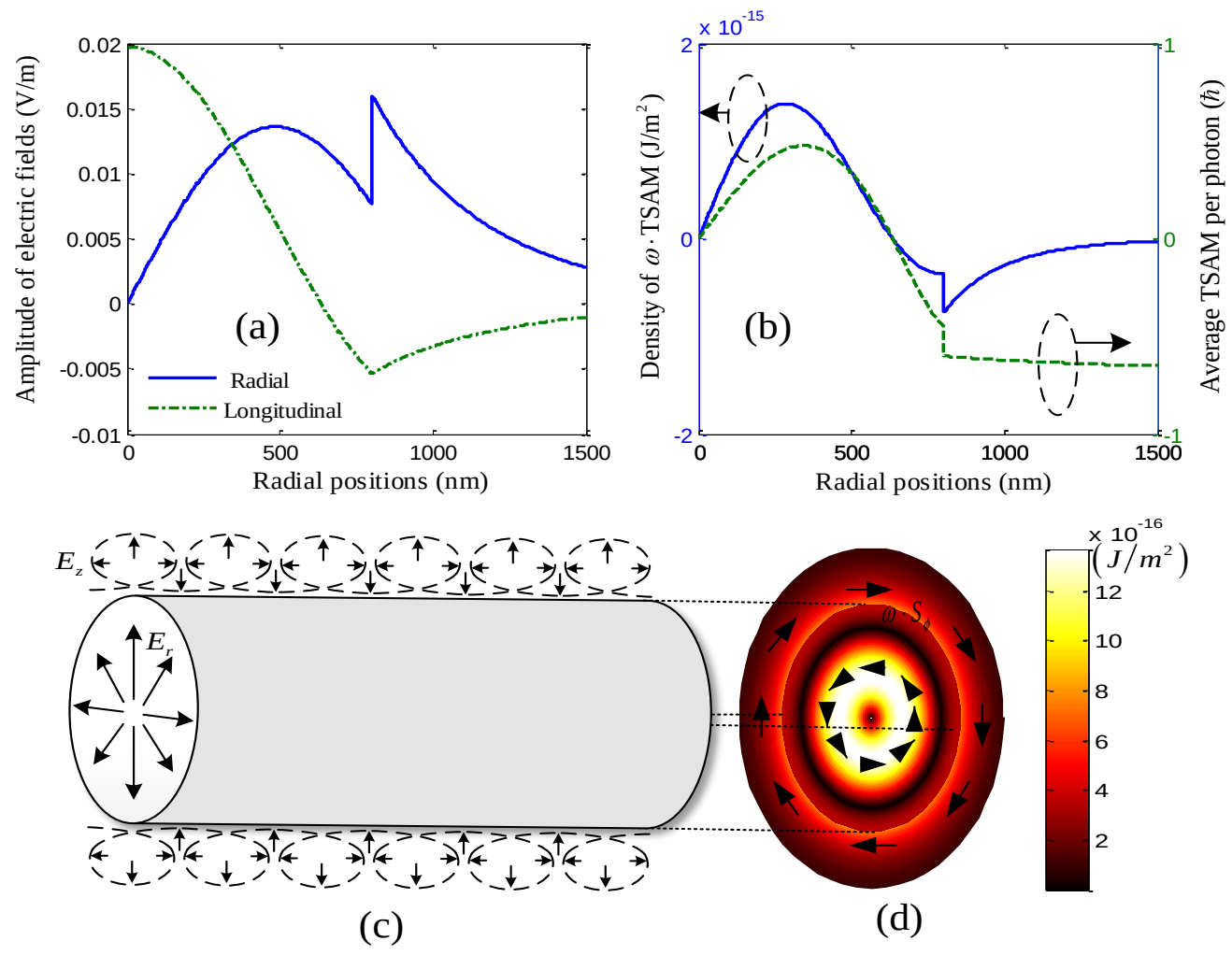

Fig. 1 (a) The radial and longitudinal electric field distribution of $TM_{01}$ mode along radial positions of nanofiber with a radius of $a = 800nm$. (b) The density of $\omega \cdot$TSAM and average TSAM per photon of $TM_{01}$ mode along radial positions. (c) The transverse helicity of evanescent field outside the nanofiber. (d) The density of $\omega \cdot$TSAM and spin direction distribution inside and outside the nanofiber. The power of optical field is 1 W.

For HE/EH modes, the radial and azimuthal components of TSAM can be respectively given by

$$S_r = \begin{cases} -\frac{A^2 a\beta\varepsilon_0}{4\omega u} J_n\left(\frac{u}{a}r\right) \begin{bmatrix} \frac{(1-s_0)}{2} J_{n-1}\left(\frac{u}{a}r\right) \\ +\frac{(1+s_0)}{2} J_{n+1}\left(\frac{u}{a}r\right) \end{bmatrix} \sin 2(n\phi + \varphi_0) & r < a \\ -\frac{A^2 a\beta\varepsilon_0 J_n^2(u)}{4\omega w K_n^2(w)} K_n\left(\frac{w}{a}r\right) \begin{bmatrix} \frac{(1-s_0)}{2} K_{n-1}\left(\frac{w}{a}r\right) \\ -\frac{(1+s_0)}{2} K_{n+1}\left(\frac{w}{a}r\right) \end{bmatrix} \sin 2(n\phi + \varphi_0) & r \geq a \end{cases} \tag{23}$$

$$S_\phi = \begin{cases} -\frac{A^2 a\beta\varepsilon_0}{2\omega u} J_n\left(\frac{u}{a}r\right) \begin{bmatrix} \frac{(1-s_0)}{2} J_{n-1}\left(\frac{u}{a}r\right) \\ -\frac{(1+s_0)}{2} J_{n+1}\left(\frac{u}{a}r\right) \end{bmatrix} \cos^2(n\phi + \varphi_0) & r < a \\ -\frac{A^2 a\beta\varepsilon_0 J_n^2(u)}{2\omega w K_n^2(w)} K_n\left(\frac{w}{a}r\right) \begin{bmatrix} \frac{(1-s_0)}{2} K_{n-1}\left(\frac{w}{a}r\right) \\ +\frac{(1+s_0)}{2} K_{n+1}\left(\frac{w}{a}r\right) \end{bmatrix} \cos^2(n\phi + \varphi_0) & r \geq a \end{cases} \tag{24}$$

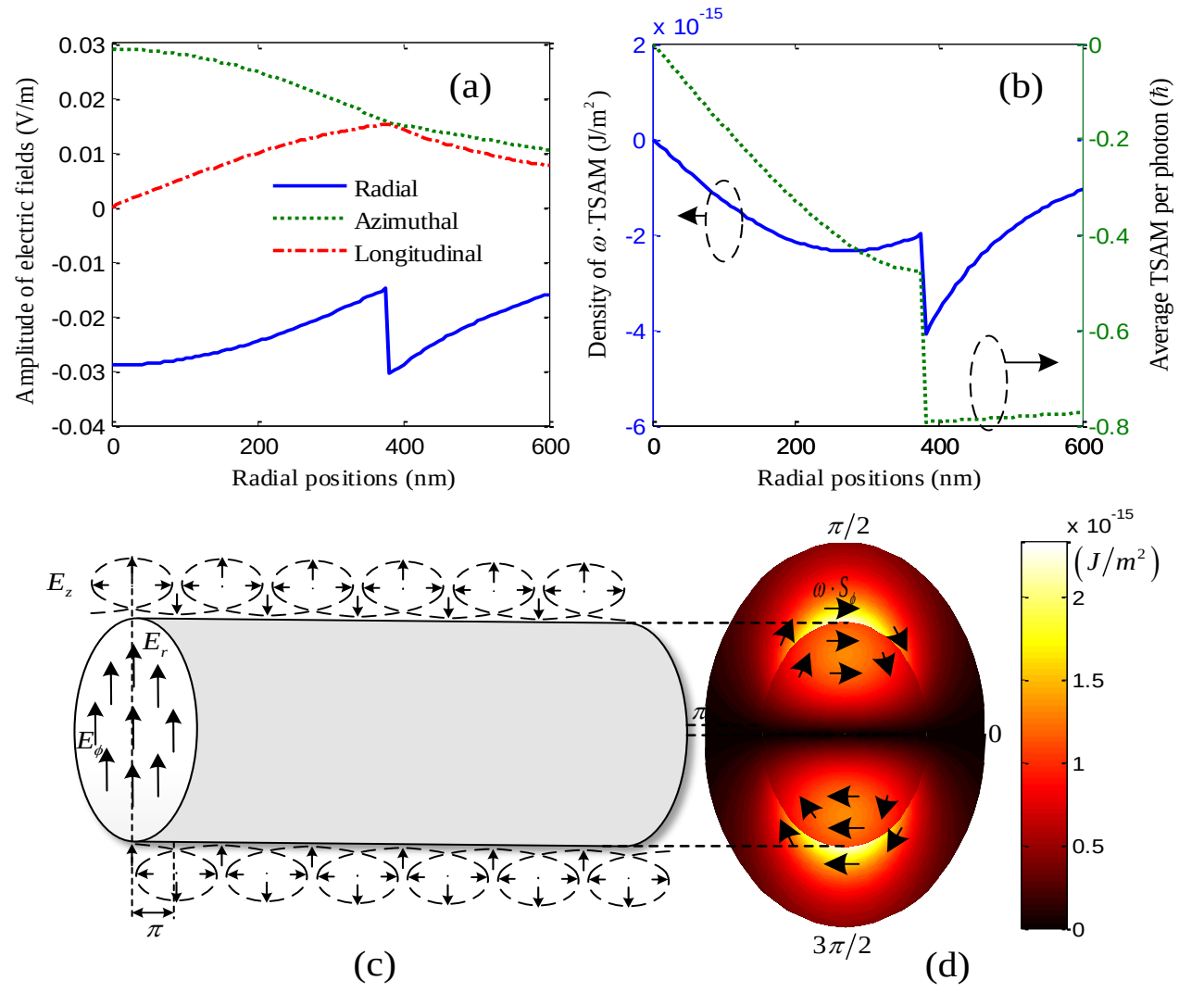

Fig. 2 (a) The radial, azimuthal and longitudinal electric field distribution of $HE_{11}$ mode along radial positions of nanofiber with a radius of $a = 380nm$. (b) The density of $\omega \cdot$TSAM and average TSAM per photon of $HE_{11}$ mode along radial positions. (c) The transverse helicity of evanescent field outside the nanofiber of y-polarized $HE_{11}$ mode. (d) The density of $\omega \cdot$TSAM and spin direction distribution inside and outside the nanofiber. The power of optical field is 1 W.

We show the transverse helicity of evanescent field outside the nanofiber in Figs. 1(c) and 2(c), and the density of $\omega\cdot$TSAM and the spin direction distribution in Figs. 1(d) and 2(d) for $TM_{01}$ and $HE_{11}$ modes, respectively. The transverse helicity consists of the transverse and longitudinal electric fields with a phase difference $\pi/2$ in the direction of propagation. It produces the TSAM of fiber-guided eigenmodes. The spin direction inside and outside fiber appears in reverse for $TM_{01}$ mode, which is due to the inverse direction of longitudinal electric field along the radial direction shown in Fig. 1(a). For the $HE_{11}$ mode, the azimuthal SAM component dominates and appears as a couple of lobes. It should be noted that the transverse helicity of $HE_{11}$ mode on the top of nanofiber has a phase offset of $\pi$ relative to that on the bottom along the propagation direction of light, whereas the $TM_{01}$ mode has not phase offset around the nanofiber. For the higher order modes, such as $HE_{21}$ and $EH_{11}$ modes, as shown in Fig. 3, the TSAM distribution inside and outside the nanofiber is characterized by two couples and one couple of lobes, respectively, which is aligned with the modal azimuthal order. Each lobe has different phase locations around the nanofiber.

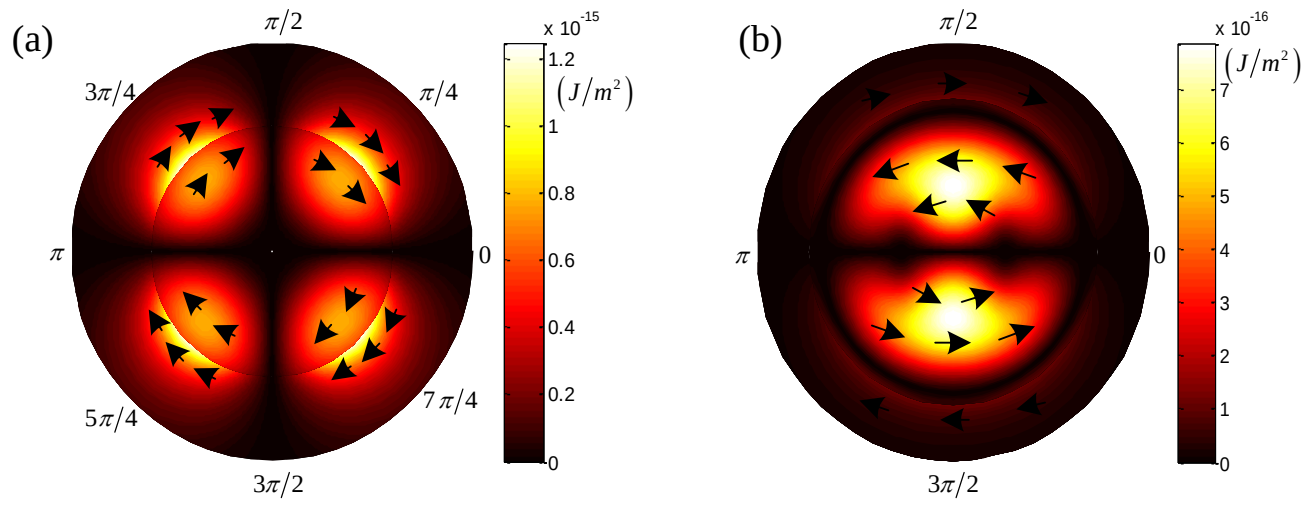

Fig. 3 The density of $\omega\cdot$TSAM and spin direction distribution of (a) $HE_{21}$ and (b) $EH_{11}$ modes in nanofibers.

Considering that the radial electric field sharply increases outside the interface between the nanofiber and surrounding, as the evanescent wave shown in Figs. 1(a) and 2(a), the TSAM would become large abruptly at this interface, as shown at the left y-axis in Figs. 1(b) and 2(b), respectively. Based on Eqs. (23) and (25), the surface TSAM $S_\phi(a)$ is highly dependent on the fiber size. Before the discussion about the surface TSAM, we give the expression of total power of each eigenmodes, which can be obtained by the integral of Eq. (1) without the contribution of magnetic field energy, $\langle P\rangle = c\langle W\rangle / n_{1,2}^2$, where $n_{1,2}$ corresponds to $n_1$ and $n_2$, respectively, when integrating over the region of fiber core and surrounding. For $TM_{01}$ mode, inserting Eqs. (11)-(14) into this expression and taking the total power of 1 W, the amplitude coefficient $A$ can be deduced as

$$A = \frac{1}{a^2}\sqrt{\frac{2Z_0}{\pi}}\left[\begin{array}{l} n_1^2 k_0^2\left(1-\dfrac{n_1^2}{n_2^2}\right)\dfrac{J_1^2(u)}{u^2}+ \\ \dfrac{\beta^2 J_0^2(u) K_2(w)}{w^2 K_0(w)} - \dfrac{\beta^2 J_0(u) J_2(u)}{u^2} \end{array}\right]^{-1/2}, \quad (25)$$

Note that in the process of deduction, we adopt the relational expression of eigenmode equations applied with [22]. When inserting Eq. (25) into Eq. (22), we can write the surface TSAM as

$$S_\phi(a) = -\frac{k_0 J_0(u) K_1(w)}{\beta w \omega^2 a^3 K_0(w)}\left[\begin{array}{l} \dfrac{n_1^2 k_0^2}{\beta^2}\left(1-\dfrac{n_1^2}{n_2^2}\right)\dfrac{J_1^2(u)}{u^2} \\ +\dfrac{J_0^2(u) K_2(w)}{w^2 K_0(w)} - \dfrac{J_0(u) J_2(u)}{u^2} \end{array}\right]^{-1}. \quad (26)$$

For nanofiber with refractive index of $n_1 = 1.45$ and $n_2 = 1$, the theoretical analyses show that $S_\phi(a)$ has an extremum value when the ratio of fiber radius to wavelength satisfies the relation of $a/\lambda \simeq 0.53$, i.e. the first derivative $S'_\phi(a/\lambda)|_{a/\lambda\simeq 0.53} = 0$ for $TM_{01}$ mode. It is expected that the surface TSAM around nanofiber is available to some applications, such as spin-controlled unidirectional propagation, particle manipulation, fiber sensing, etc. Applying similar procedures to the HE/EH modes, we can get the extremum condition of surface TSAM, that is, $a/\lambda \simeq 0.25$ for $HE_{11}$ mode, 0.47 for $HE_{21}$ mode, 0.72 for $EH_{11}$ mode, and 0.67 for $HE_{31}$ mode, respectively. Figs. 4(a) and 4(b) show the relation between the value of surface TSAM and fiber radii for $TM_{01}$ and $HE_{11}$ modes, respectively, at the wavelength of 1550 nm. The optimal fiber size is about 800nm and 380nm for these two modes to get the maximum surface TSAM for $TM_{01}$ and $HE_{11}$ modes, respectively, which is constant with the aforementioned conclusion. This extremum condition coincides with the numerical results by F. Kalhor [24].

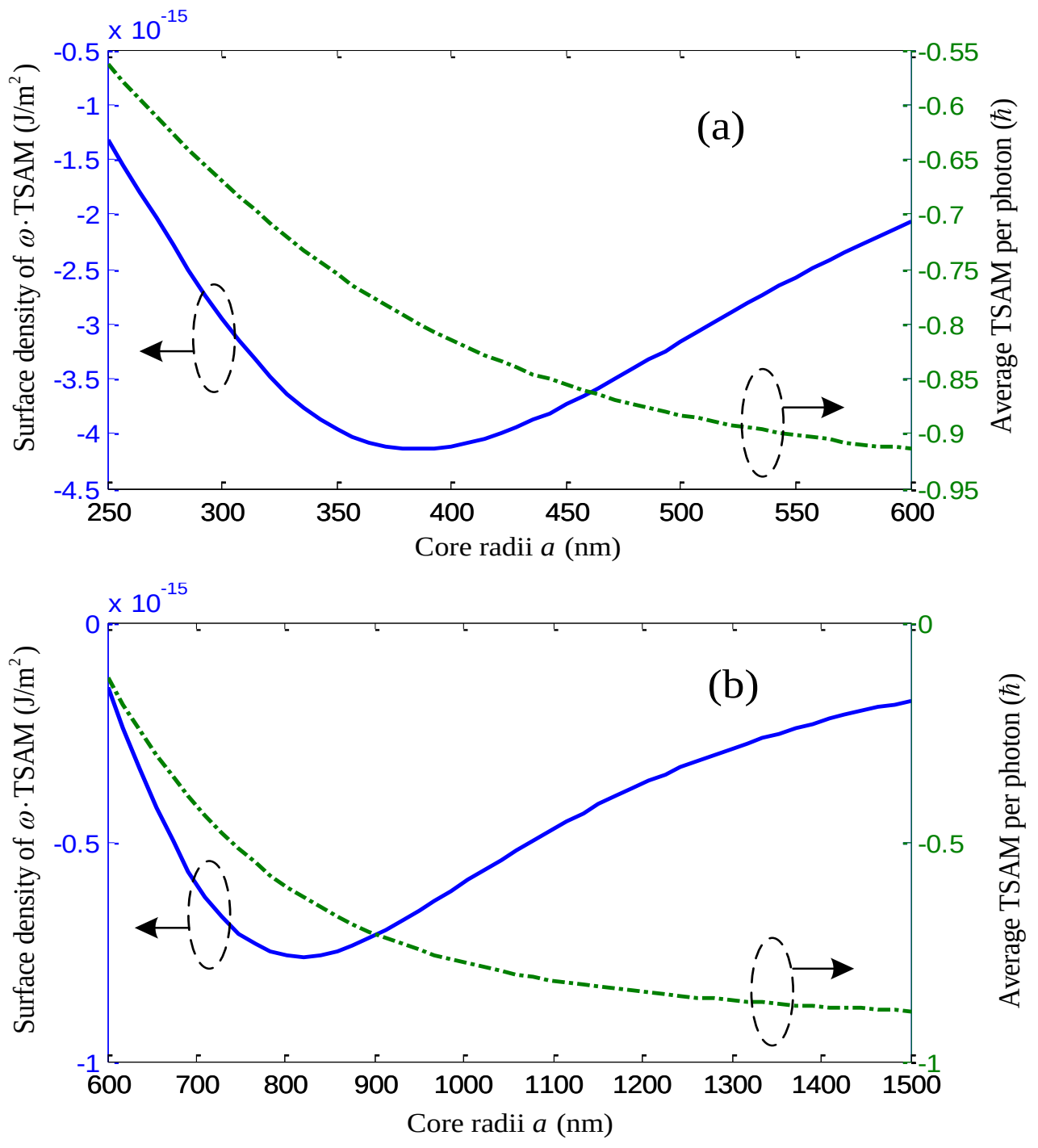

Fig. 4 The values of surface TSAM density and average TSAM per photon versus fiber core radii for (a) $HE_{11}$ (a) and (b) $TM_{01}$ modes.

The average TSAM per photon can be expressed by $\left[S_\phi/W\right]\cdot\hbar\omega$, where $\hbar = h/2\pi$, with Planck's constant $h$. The average TSAM per photon of $TM_{01}$ and $HE_{11}$ modes along the radial direction of nanofiber is shown at the right y-axis in Figs. 1(b) and 2(b), respectively. As shown in Figs. 4(a) and 4(b), the average TSAM per photon on the fiber surface reduces with the increase of the fiber core radius. The phenomenon of maximum values of TSAM densities in fiber core radius of 380nm and 800nm is ascribed to the large density of photon numbers $W/\hbar\omega$ on the fiber surface. The quantum number density per photon of TSAM has the limiting values $\pm 1/n_1^2$ inside nanofiber and $\pm 1/n_2^2$ outside nanofiber, and both the absolute values of them are less than or equal to 1.

The total integral expression of TSAM is given by $\langle S_\phi\rangle = \langle S_\phi\rangle_{\text{core}} + \langle S_\phi\rangle_{\text{sur.}}$, where $\langle S_\phi\rangle_{\text{core}} = \iint_{\text{core}} S_\phi r dr d\phi$, and $\langle S_\phi\rangle_{\text{sur.}} = \iint_{\text{sur.}} S_\phi r dr d\phi$, which stand for the integral AM in the core and surrounding of nanofiber, respectively. Note that the integral value of azimuthal component of TSAM $\langle S_r\rangle = 0$ has no contribution to the total integral TSAM. The integral form of average TSAM per photon is given by $j_s = \left[\langle S_\phi\rangle/\langle W\rangle\right]\hbar\omega$ that represents the average TSAM per photon of the total field. This average TSAM per photon has a limiting value of $j_s = \pm\hbar/n_1^2$, and $|j_s| < 1\hbar$ inside nanofiber, which is consistent with Abraham expression for the AM in a dielectric medium [25]. We calculate the integral TSAM of nanofiber versus the fiber core radii for $TM_{01}$ and $HE_{11}$ modes, as shown in Figs. 5(a) and 5(b), respectively. They include both the TSAM contributions inside the fiber core and around the surrounding. One can see that the integral average TSAM per photon around the fiber surrounding has almost the same extremum condition as the surface TSAM density.

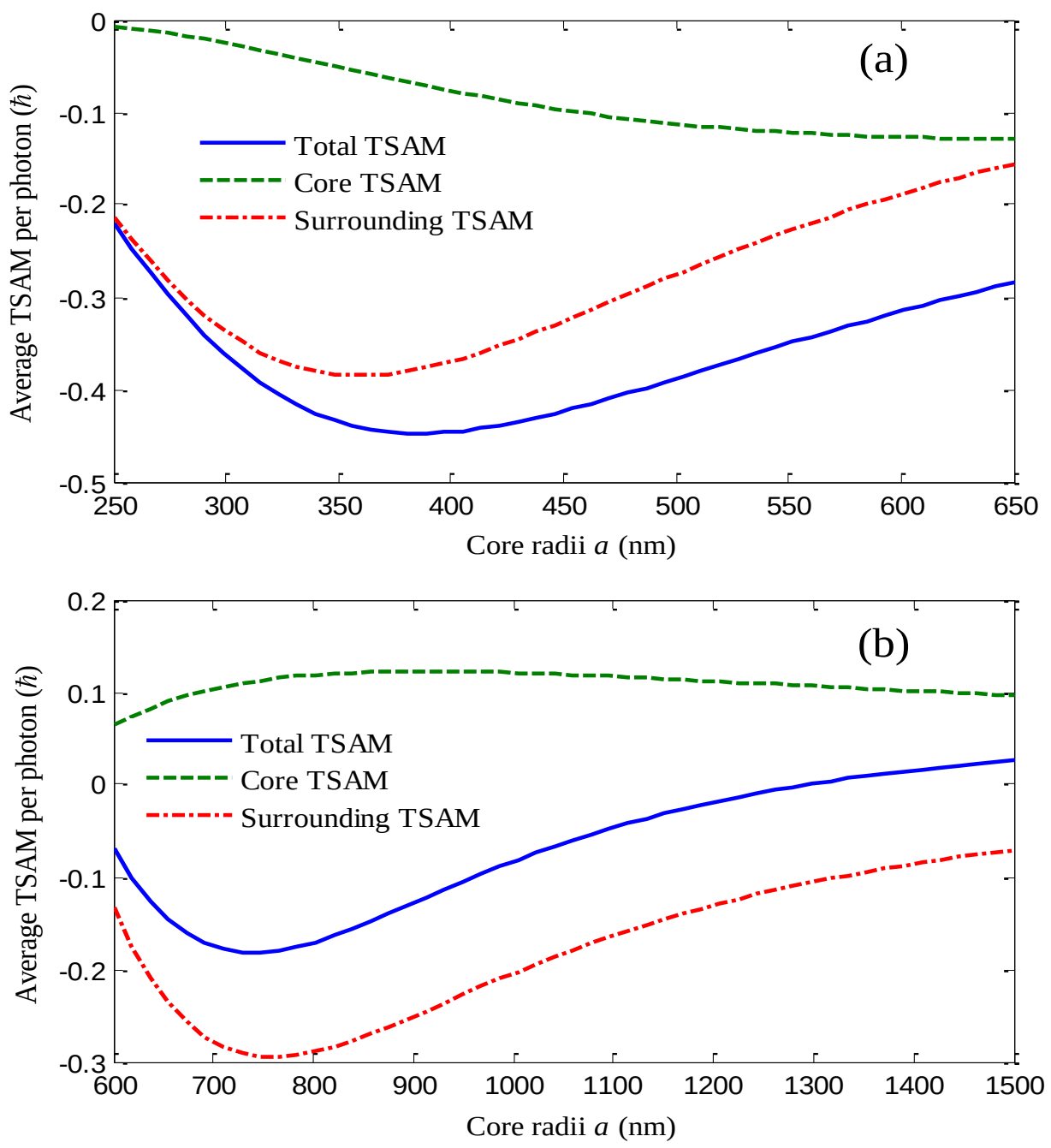

Fig. 5 The integral values of TSAM per photon in fiber core,

surrounding, and the total TSAM versus fiber core radii for (a) $HE_{11}$ and (b) $TM_{01}$ modes.

We then study the spin flows in the case of circularly polarized fiber-guided modes. As is well known, the circular polarization induces the conventional longitudinal SAM. The longitudial SAM per photon is +1ħ for the right-handed circular polarization, whereas is -1ħ for the left-handed circular polariation. For the circularly polarized $HE_{11}$ mode, the combined trajectory of all SAM components (including the classical longitudinal SAM and new transverse SAM) is characterized by a helix as shown in Fig. 6. This helical SAM flow can bring the chiral optical forces [26]. It is originated from the phase offset of the lobes of TSAM along the propagation direction as shown in Fig. 2(c). One can see that the handedness of spin flow is in accordance with that of polarization states. The spin flows along the propagation direction of light for the right-handed circularly polarized modes, however, in reverse for the left-handed circularly polarized modes. The remarkable properity of the TSAM is that the direction of TSAM is just determined by the propagation direction of light, regardless of the polarization handedness of fiber eigenmodes. It is associated with the phenomenon of spin-momentum locking (transverse spin direction is locked to the direction of linear momentum of light). It belongs to the intrinsic property of TSAM, which makes sense that it can be used to spin-controlled unidirectional propagation of light via scattering by nanoparticle [12]. However, for the special $TM_{01}$ mode with radial polarization, its spin trajectory is just a closed-loop, as shown in Figs. 7(a) and 7(b), because of no phase offsets of TSAM shown in Figs. 1(c) and 1(d), as well as no longitudinal SAM. For higher-order circularly polarized HE/EH modes, for example, the $HE_{21}$ mode, apart from the feature of carrying the usual OAM, its spin flows have 2-fold helical trajectories due to the phase offset shown in Fig. 3(a). In general, the fold number of spin flowing trajectories of the circularly polarized fiber-guided higher-order modes corresponds to the number of azimuthal order $n$ of these modes.

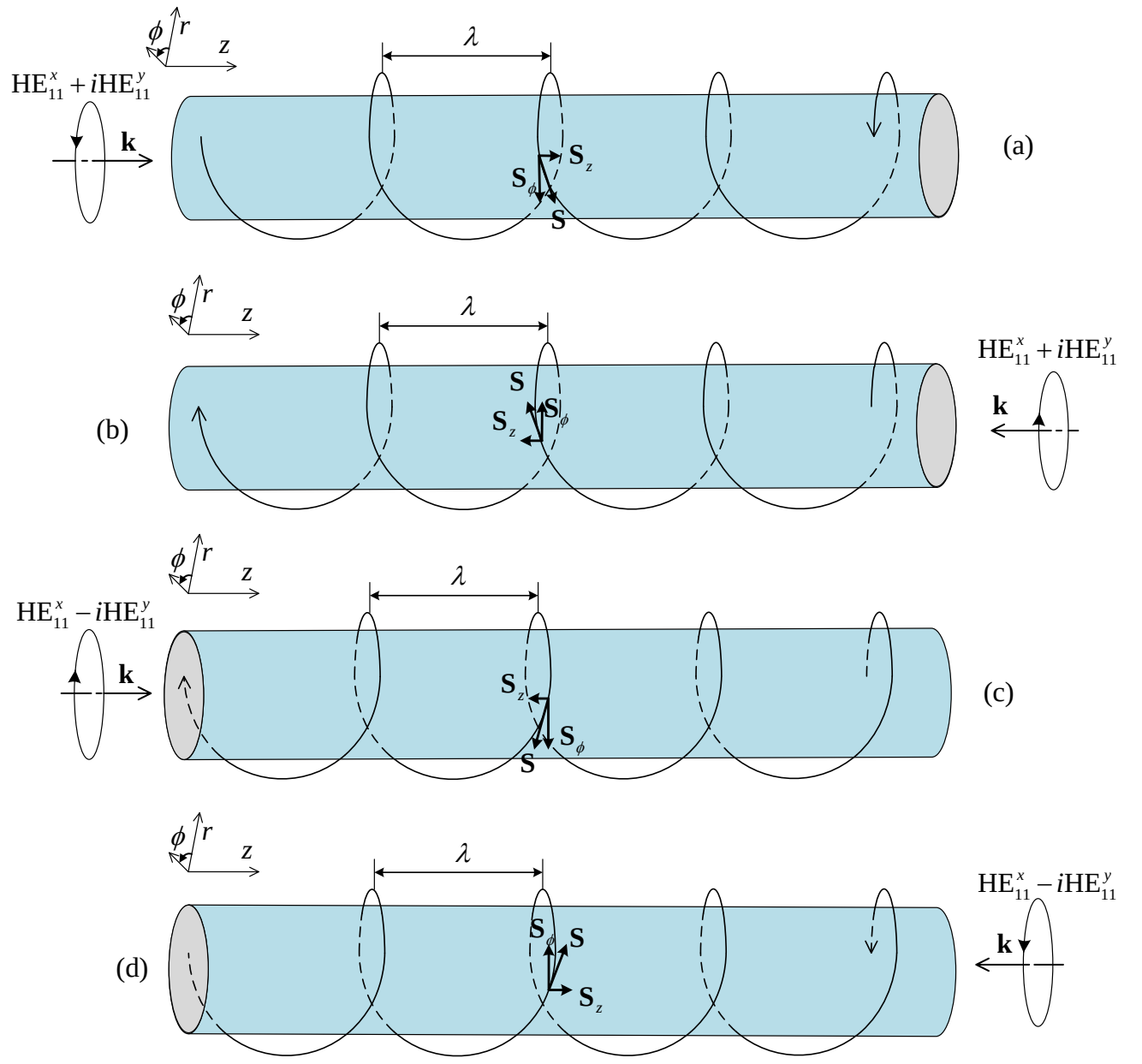

Fig. 6 Spin flows on nanofiber surface for (a) right-handed circularly polarized $HE_{11}$ mode along the $+z$ direction, (b) right-handed circularly polarized $HE_{11}$ mode along the $-z$ direction, (c) left-handed circularly polarized $HE_{11}$ mode along the $+z$ direction, and (d) left-handed circularly polarized $HE_{11}$ mode along the $-z$ direction.

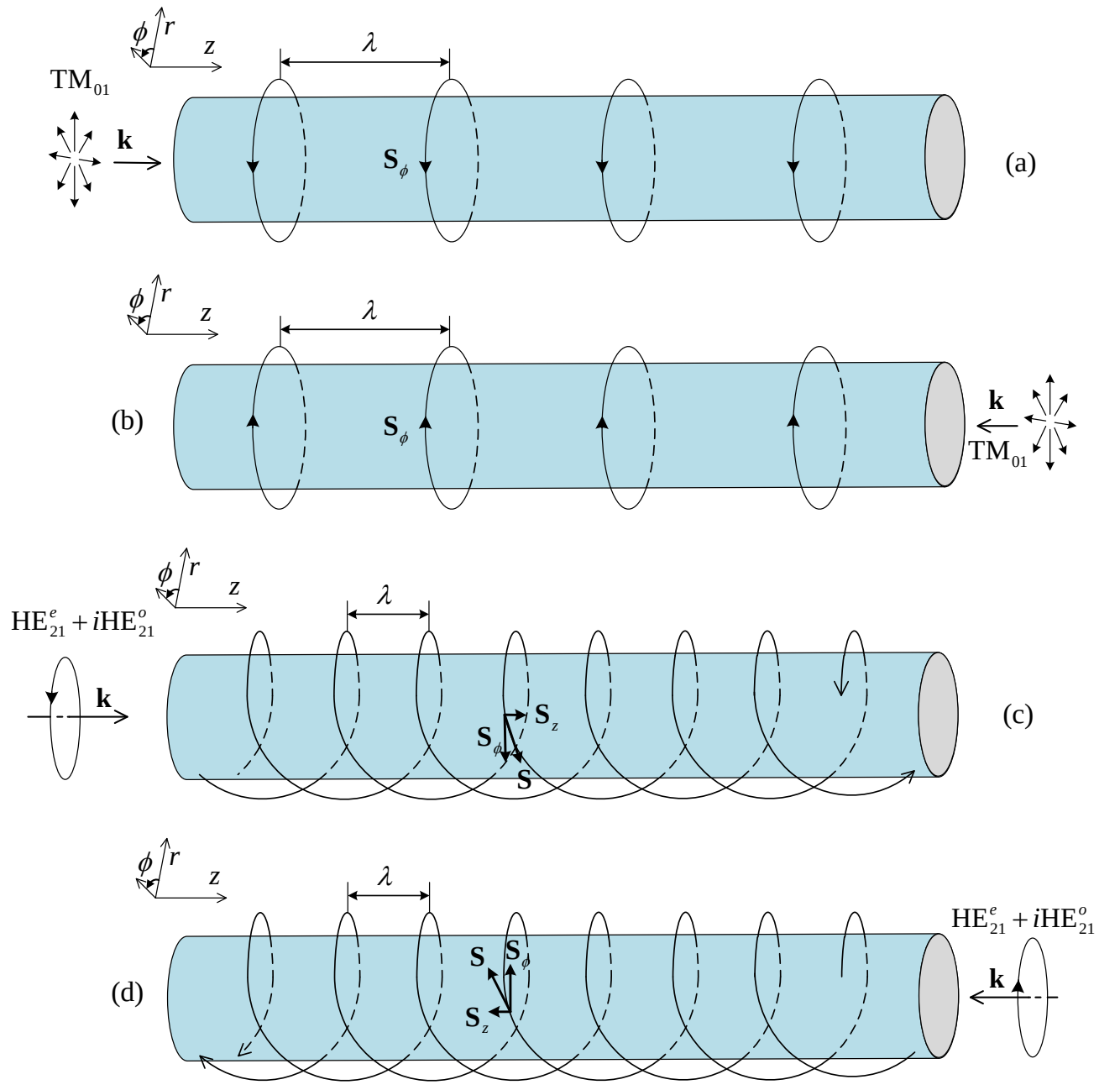

Fig. 7 Spin flows on nanofiber surface for (a) $TM_{01}$ mode along the $+z$ direction, (b) $TM_{01}$ mode along the $-z$ direction, (c) right-handed circularly polarized $HE_{21}$ modes along the $+z$ direction, and (d) right-handed circularly polarized $HE_{21}$ modes

along the $-z$ direction.

We further study the mechanical effects originated from TSAM around nanofiber, including lateral force on chiral particle from the spin-momentum locking, and transverse optical torque on dipole Rayleigh particle via the electric-dipole coupling. The lateral force is caused by this phenomenon associated with the intrinsic quantum spin Hall effect of light [10, 11, 13]. For a small particle that has a sufficiently small radius relative to the wavelength $\lambda$, the electric-electric polarizability can be simplified by [7]

$$\alpha_{ee} \simeq \frac{1}{k^3} \frac{\varepsilon_r^0 \left(\varepsilon_p - \varepsilon_r^0\right)\left(ka_0\right)^3}{\varepsilon_p + 2\varepsilon_r^0}, \tag{27}$$

where $\varepsilon_p$ denotes complex relative permittivity of material of particle and $\varepsilon_r^0$ is the relative permittivity in the vacuum, $a_0$ is the radius of a small equivalent spherical particle. In the numerical calculation, we take $\varepsilon_p = 3 + 0.28 \cdot i$, $\varepsilon_r^0 = 1$, and $a_0 = 10\text{nm}$, and meanwhile, we assume that the particle also has chirality property, of which the chirality factor is taken as $\alpha_{em} = 1\times 10^{-26}$, liken as calculation in the references [11]. The particle is 50nm away from fiber surface, and interacted with surrounding TSAM from $HE_{11}$, $TM_{01}$, and $HE_{21}$ modes, respectively. The radius of nanofibers supporting $HE_{11}$, $TM_{01}$, and $HE_{21}$ modes is taken as 380nm, 800nm, and 720nm, respectively. The calculation results are shown in Fig. 6 when changing location of particle at different azimuthal positions around nanofibers. Fig. 8(a) shows the lateral forces resulted from chiral property of particle, which can arrive the order of magnitude of pN, and Fig. 8(b) shows the transverse optical torques. The $HE_{11}$ mode produces the largest mechanical effect, compared with other higher order modes. The values of lateral forces and transverse optical torques are proportional to the density of TSAM, and the variation period around one circle doubles to the mode order $n$ of fiber eigenmodes.

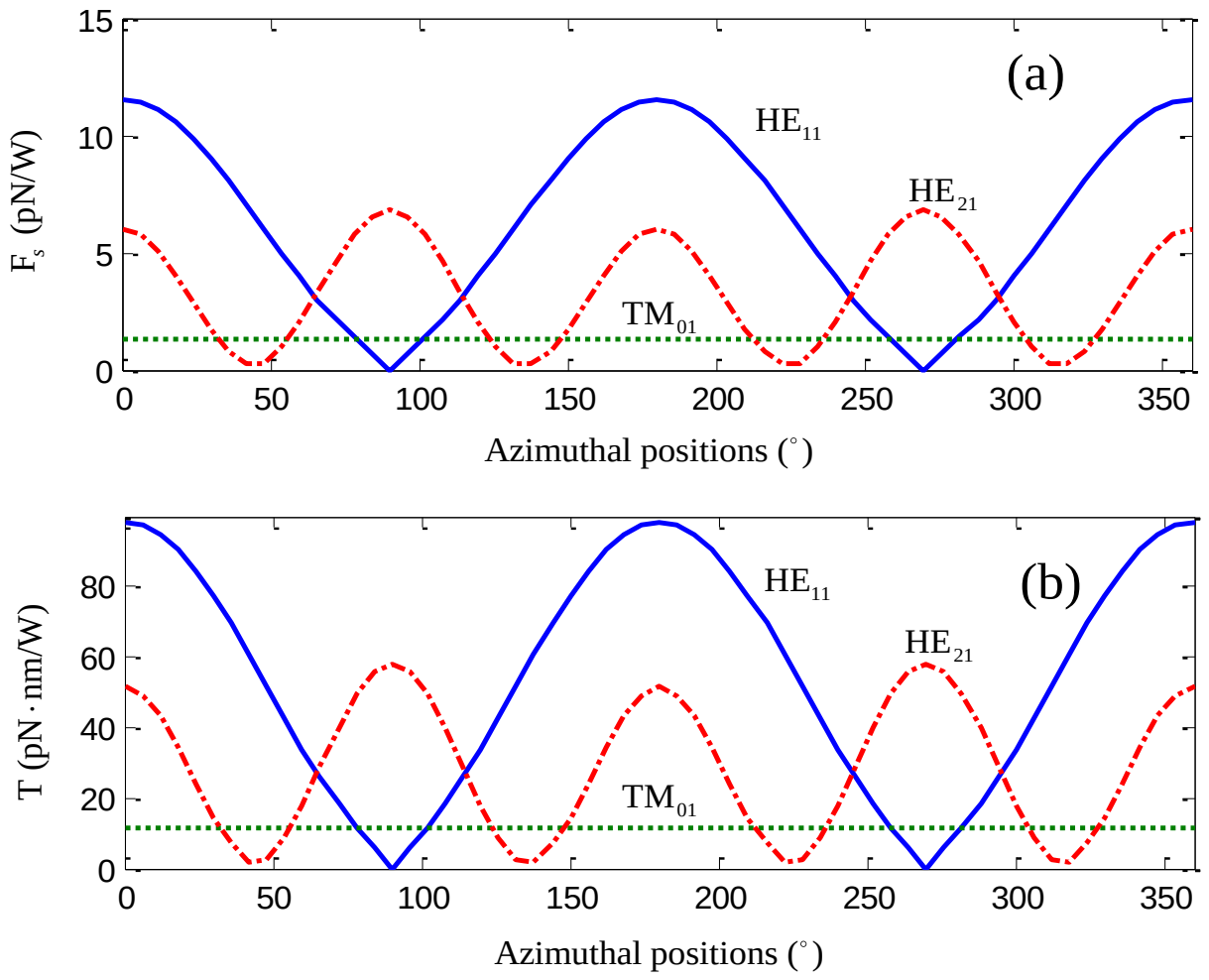

Fig. 8 (a) Lateral forces and (b) transverse optical torques of $HE_{11}$, $TM_{01}$, and $HE_{21}$ modes acting on small particle at different azimuthal positions. The equivalent spherical particle with radius $a_0 = 10\text{nm}$, $\varepsilon_p = 3 + 0.28 \cdot i$ and $\alpha_{em} = 1\times 10^{-26}$ is 50 nm away from the nanofiber surface.

In conclusion, we have comprehensively investigated the TSAM and spin flows of fiber eigenmodes, as well as its mechanical effects on small particles. We have deduced analytical expressions of the intrinsic TSAM, and given the extremum condition of surface TSAM on nanofiber for each fiber-guide eigenmodes. For instance, for the fundamental mode, we can get the maximum surface TSAM in the case that the fiber radius is one-fourth of the wavelength. It makes sense that the surface TSAM around nanofiber is available to some applications, such as spin-controlled unidirectional propagation or particle manipulation. Furthermore, we have presented the spin flows of circularly polarized fiber-guided modes, and numerically calculated the lateral force on chiral particle from spin-momentum locking, as well as the transverse optical torque on dipole Rayleigh particle via the electric-dipole coupling. We believe that our investigation and revelation may be advantage to potential utilization and exploitation of fiber modes and their TSAM in the realms of fiber-based classical and quantum optical implementations, such as all-optical signal processing, particle manipulation, fiber sensing, etc.

**Acknowledgment**

This work was supported by the National Program for Support of Top-Notch Young Professionals, National Basic Research Program of China (973 Program) under grant 2014CB340004, the National Natural Science Foundation of China (NSFC) under grants 11574001, 11274131 and 61222502, and the Program for New Century Excellent Talents in University (NCET-11-0182)

**References:**

1. L. Allen, S. M. Barnett, and M. J. Padgett. Optical Angular Momentum (IoP Publishing, 2003); D. L. Andrews and M. Babiker, etc., The Angular Momentum of Light (Cambridge University Press, Cambridge, 2013).

2. K. Y. Bliokh, A. Y. Bekshaev, and F. Nori, “Dual electromagnetism: helicity, spin, momentum, and angular momentum”, New J. Phys. 15, 033026 (2013).

3. D. E. Soper, Classical Field Theory (Wiley, New York, 1976); L. D. Landau and E. M. Lifshitz, Mechanics, 3rd Ed. (Butterworth-Heinmann, Oxford, 1976).

4. A. S. Desyatnikov, Y. S. Kivshar, and L. Torner, “Optical vortices and vortex solitons”, Prog. Opt. 47, 291 (2005); L. Allen, M. W. Beijersbergen, R. J. C. Spreeuw, and J. P. Woerdman, “Orbital angular momentum of light and the transformation of Laguerre-Gaussian laser modes” Phys. Rev. A, 45, 8185 (1992).

5. C. Cohen-Tannoudji, J. Dupont-Roc, and G. Grynberg. Atom-Photon Interactions (Wiley-VCH, 2004).

6. A. Y. Bekshaev, “Subwavelength particles in an inhomogeneous light field: optical forces associated with the spin and orbital energy flows” J. Opt. 15. 044004 (2013); I. Brevik. “Experiments in phenomenological electrodynamics and the electromagnetic energy-momentum tensor,” Phys. Rep. 52, 133 (1979); D. G. Grier, “A revolution in optical manipulation” Nature, 424, 810 (2003); M. Padgett, R. Bowman, “Tweezers with a twist,” Nat. Photonics 5, 343 (2011).

7. A. Aiello, N. Lindlein, C. Marquardt, and G. Leuchs, “Transverse angular momentum and geometric spin Hall effect of light”, Phys. Rev. Lett. 103, 100401 (2009); K. Y. Bliokh, A. Y. Bekshaev, F. Nori, “Extraordinary momentum and spin in evanescent waves”, Nat. Commun. 5, 3300 (2014); K. Y. Bliokh and F. Nori, “Transverse spin of a surface polariton, ” Phys. Rev. A, 85, 061801 (2012); Z. Wang, B. Chen, R. Wang, S. Shi, Q Qiu, X. Li, “New investigations on the transverse spin of structured optical fields ”, arXiv, (2016).

8. A. Y. Bekshaev, K. Y. Bliokh, and F. Nori, “Transverse spin and momentum in two-wave interference”, Phys. Rev. X, 5, 011039 (2015); K. Y. Bliokh, A. Y. Bekshaev, F. Nori, “Extraordinary momentum and spin in evanescent waves”, Nat. Commun. 5, 3300 (2014).

9. R. Mathevet and G.L.J. A. Rikken, “Magnetic circular dichroism as a local probe of the polarization of a focused Gaussian beam”, Opt. Materials Express 4, 2574 (2014); M. Neugebauer, T. Bauer, A. Aiello, and P. Banzer, “Measuring the transverse spin density of light”, Phys. Rev. Lett. 114, 063901 (2015).

10. K. Y. Bliokh, and F. Nori, “Transverse and longitudinal angular momenta of light,” Phys. Rep. 4, 592(2015). M. H. Alizadeh, and B. M. Reinhard, “Emergence of transverse spin in optical modes of semiconductor nanowires,” Opt. Express, 24, 8471 (2016).

11. F. Kalhor, T. Thundat, and Z. Jacob, “Universal spin-momentum locked optical forces, ” App. Phys. Lett. 108, 061102 (2016); S. B. Wang and C. T. Chan, “Lateral optical force on chiral particles near a surface, ” Nat. Commun. 5, 3307 (2014); A. Hayat, J. P. B. Mueller, and F. Capasso, “Lateral chirality-sorting optical forces, ” Proc. Natl. Acad. Sci. 112, 13190 (2015).

12. J. Petersen, J. Volz, and A. Rauschenbeutel, “Chiral nanophotonic waveguide interface based on spin-orbit interaction of light”, Science, 346, 625(2014); D. O' Connor, P. Ginzburg, G. A. Wurtz, A. V. Zayats, “Spin-orbit coupling in surface plasmon scattering by nanostructures”, Nature Commun. 5, 5327 (2014); B. le Feber, N. Rotenberg, and L. Kuipers, “Nanophotonic control of circular dipole emission: towards a scalable solid-state to flying-qubits interface”, Nature Commun. 6, 6695 (2015).

13. K.Y. Bliokh and F. Nori, “Quantum spin Hall effect of

light", Science 348, 1448 (2015).
14. C. Sayrin, C. Junge, R. Mitsch, and B. Albrecht, "Nanophotonic Optical Isolator Controlled by the Internal State of Cold Atoms", Phys. Rev. X 5, 041036 (2015). R. J. Coles, D. M. Price, J. E. Dixon, B. Royall, E. Clarke, A. M. Fox, P. Kok, "Chirality of nanophotonic waveguide with embedded quantum emitter for unidirectional spin transfer," Nature Commun. 7, 11183 (2016).
15. S. Golowich, "Asymptotic theory of strong spin–orbit coupling in optical fiber", Opt. Lett. 39, 92 (2014); F. L.e Kien, J. Q. Liang, K. Hakuta, and V. I. Balykin, "Field intensity distributions and polarizationorientations in a vacuum-clad subwavelength-diameter optical fiber," Opt. Commun. 242, 445(2004).
16. F. L. Kien, V. I. Balykin, and K. Hakuta, "Angular momentum of light in an optical nanofiber", Phys. Rev. A, 73, 053823 (2006); K. Y. Bliokh, F. J. Rodríguez-Fortuño, F. Nori, and A. V. Zayats, "Spin-orbit interactions of light", Nat. Photon. 9, 796(2015).
17. H. Murata, Handbook of Optical Fibers and Cables 2nd edn (Marcel Dekker, New York, 1996); D. K. Mynbaev, amd L. L. Scheiner, Fiber-Optic Communications Technology (Prentice Hall, New York,2001); L. Tong, R. R. Gattass, J. B. Ashcom, S. He, J. Lou, M.Shen, I. Maxwell, E. Mazur, "Subwavelength-diameter silica wires for low-loss optical wave guiding," Nature 426, 816 (2003);
18. K. Y. Bliokh and F. Nori, "Transverse spin of a surface polariton, " Phys. Rev. A, 85, 061801 (2012); K. Y. Bliokh, and F. Nori, "Characterizing optical chirality", Phys. Rev. A, 83, 021803 (2011);
19. Y. Tang, and A. E. Cohen, "Optical chirality and its interaction with matter," Phys. Rev. Lett. 104, 163901 (2010).
20. Y. Tang and A. E. Cohen, "Enhanced Enantioselectivity in Excitation of Chiral Molecules by Superchiral Light", Science, 332, 333(2014).
21. A. Lakhtakia, Selected articles on Natural Optical Activity (SPIE Optical Engineering Press, Bellingham, 1990); I. V. Lindell, A. H. Sihvola, S. A. Tretyakov, A. J. Viitanen. Electromagnetic Waves in Chiral and Bi-Isotropic Media (Artech House, 1994).
22. K. Okamoto, Fundamentals of Optical Waveguides (Elsevier Academic Press, 2006), Chap. 3.
23. Y. I. Salamin, "Electron acceleration from rest in vacuum by an axicon Gaussian laser beam," Phys. Rev. A 73, 043402 (2006); Q. Zhan, "Trapping metallic Rayleigh particles with radial polarization," Opt. Express 12, 3377–3382 (2004); M. Meier, V. Romano, and T. Feurer, "Material processing with pulsed radially and azimuthally polarized laser radiation," Appl. Phys. A 86, 329–334 (2007).
24. F. Kalhor, T. Thundat, and Z. Jacob, "Universal spin-momentum locked optical forces, " App. Phys. Lett. 108, 061102 (2016).
25. F. L. Kien, V. I. Balykin, and K. Hakuta, "Angular momentum of light in an optical nanofiber", Phys. Rev. A, 73, 053823 (2006); M. Padgett, S. M. Barnett, and R. Loudon, "The angular momentum of light inside a dielectric ", J. Mod. Opt. 50, 1555 (2003).
26. M. H. Alizadeh, and B. M. Reinhard, "Bringing chiral optical forces to dominance with optical nanofibers," arXiv:1605.06008v1 [physics.optics].